\documentclass[runningheads]{llncs}

\usepackage{url}
\usepackage{graphicx}
\usepackage{booktabs}
\usepackage{makecell}
\usepackage{listings}
\usepackage{color}
\usepackage{authblk}

\usepackage{hyperref}
\hypersetup{
    colorlinks=true,
    linkcolor=black,
    citecolor=black,
    filecolor=black,
    urlcolor=black,
}

\newcommand{\ignore}[1]{}

\begin{document}
\title{Lucene for Approximate Nearest-Neighbors Search on Arbitrary Dense Vectors}

\titlerunning{Lucene for ANN Search on Arbitrary Dense Vectors}
\authorrunning{T. Teofili and J. Lin}

\author{Tommaso Teofili\inst{1} and Jimmy Lin\inst{2}}

\institute{Adobe \and University of Waterloo}

\maketitle

\begin{abstract}

We demonstrate three approaches for adapting the open-source Lucene search library to perform approximate nearest-neighbor search on arbitrary dense vectors, using similarity search on word embeddings as a case study.
At its core, Lucene is built around inverted indexes of a document collection's (sparse) term--document matrix, which is incompatible with the lower-dimensional dense vectors that are common in deep learning applications.
We evaluate three techniques to overcome these challenges that can all be natively integrated into Lucene:\ the creation of documents populated with fake words, LSH applied to lexical realizations of dense vectors, and k--d trees coupled with dimensionality reduction.
Experiments show that the ``fake words'' approach represents the best balance between effectiveness and efficiency.
These techniques are integrated into the Anserini open-source toolkit and made available to the community.


\end{abstract}

\section{Introduction}

There is no doubt that the open-source Lucene search library is the most widely-adopted solution for developers seeking to build production search applications.
While it is true that commercial search engine companies such as Google and Bing deploy custom infrastructure, most organizations today---including Apple, Bloomberg, Reddit, Twitter, and Wikipedia---all use Lucene, typically via Solr or Elasticsearch.
There is, however, one important missing feature in Lucene:\ the ability to perform nearest-neighbor search on arbitrary vectors.
Our work addresses this gap.

With the advent of deep learning and neural approaches to both natural language processing and information retrieval, this is a major shortcoming of Lucene.
Such a feature is needed, for example, to look up similar words based on word embeddings.
Additionally, researchers have been developing neural models that {\it directly} attempt to minimize some simple metric (e.g., cosine distance) between ``queries'' and ``documents'' for retrieval tasks~\cite{Henderson:1705.00652:2017,Zamani:2018:NRN:3269206.3271800,Ji:2019:EIN:3308558.3313576}, which require fast nearest-neighbor search on collections of arbitrary vectors.

At its core, Lucene is built around inverted indexes of a document collection's term--document matrix.
Since the feature space comprises the vocabulary, the vectors are very sparse.
In contrast, deep learning applications mostly use dense vectors, typically only a few hundred dimensions (e.g., word embeddings), which are not directly compatible with inverted indexes.
Similarity search typically requires an entirely different set of techniques, most often based on some variant of locality-sensitive hashing~\cite{gionis1999similarity,Andoni_Indyk_CACM2008}.
As a result of these fundamental differences, systems that require both capabilities---for example, a ranking architecture that uses inverted indexes for candidate generation followed by a model exploiting vector similarity---typically cobble together heterogeneous components.

What if this wasn't necessary?
Our demonstration explores techniques for performing approximate nearest-neighbor search on dense vectors {\it directly in Lucene}.
We examine three approaches to the specific problem of retrieving similar word embedding vectors.
Experimental results show that the ``fake words'' approach provides reasonable effectiveness and efficiency.
Although, admittedly, our solutions lack elegance, they can be directly implemented in Lucene without any external dependencies.

\section{Methods}

We examine three techniques for implementing approximate nearest-neighbor search on dense vectors within Lucene, outlined below:

\smallskip \noindent {\bf ``Fake words''.} We implement the approach described in Amato et al.~\cite{Amato_etal_2016}, which encodes the features of a vector as a number of ``fake'' terms proportional to the feature value according to the following scheme:\
Given a vector $w = (w_1, . . . , w_m)$, each feature $w_i$ is associated with a unique alphanumeric term $\tau_i$ so that the document corresponding to the vector $w$ is represented by fake words generated by $\cup_{i=1}^{m}\cup_{j=1}^{\lfloor Q \cdot w_i \rfloor}\tau_i$, where $Q>1$ is a quantization factor.
Thus, the fake words encoding maintains direct proportionality between the float value of a feature and the term frequency of the corresponding fake index term.
Feature-level matching for retrieval is achieved by matching on these fake words with scores computed by Lucene's \texttt{ClassicSimilarity} (a tf-idf variant).
Finally, for this approach to be effective, vector inner products have to be equivalent to cosine similarity, which can be achieved by normalizing the vectors to unit length.

\smallskip \noindent {\bf ``Lexical LSH''.} 
We implement an approach that \textit{lexically} quantizes vector components for easy indexing and search in Lucene using LSH.
Given a vector $w = (w_1, . . . , w_m)$, each feature $w_i$ is rounded to the first decimal place and tagged with its feature index $i$.
For example, $w = \{0.12, 0.43, 0.74\}$ is realized as the tokens \texttt{1\_0.1}, \texttt{$2\_0.4$}, and \texttt{$3\_0.7$}.
Optionally, tokens are aggregated into $n$-grams and finally passed to an LSH function (already implemented in Lucene as \texttt{MinHashFilter}) to hash the tokens (or $n$-grams) into a configurable number of buckets; see Gionis et al.~\cite{gionis1999similarity}.
Thus, the vector $w$ is represented as a set of LSH-generated text signatures for \textit{tagged} and \textit{quantized} feature $n$-grams.

\smallskip \noindent {\bf k--d trees.}
We leverage Lucene's existing capability to index $n$-dimensional points comprised of floating point values, which is based on k--d trees, to perform nearest-neighbor search.
The Lucene implementation currently suffers from the limitation of being able to handle at most eight dimensions, and therefore k--d trees can only be used after dimensionality reduction.
In order to accomplish this, we use either PCA~\cite{wold1987principal} or post-processing from Mu et al.~\cite{mu2017all} combined with PCA, as in Raunak~\cite{raunak2017simple}.

\section{Experiments}

We choose nearest-neighbor search on dense word embedding vectors as our representative task for evaluation.
Specifically, we considered word2vec~\cite{mikolov2013distributed}, trained on a GoogleNews corpus, and GloVe~\cite{pennington2014glove}, trained on a Twitter corpus, both having 300 dimensional vectors.

All techniques discussed in the previous section are implemented in the Anserini toolkit~\cite{Yang_etal_JDIQ2018}\footnote{\url{http://anserini.io/}} and are released along with this demonstration.
For the ``fake words'' and ``lexical LSH'' approaches, there are a number of parameters that control effectiveness--efficiency tradeoffs, which we tune specifically for word2vec and GloVe.
We argue that usual notions of segregating training and test sets are not applicable here, because the word embeddings are static and provided in advance---and thus there is no reason why a researcher wouldn't optimally tune parameters for the specific corpus.

One more implementation detail is worth mentioning:\ for the ``fake words'' and ``lexical LSH'' approaches, we observe a large number of terms that are generated at indexing time, which significantly reduces search performance.
To combat this, we filter highly-frequent terms at search time.
Once again, this filtering threshold is tuned per collection, and we observe that this technique gives us both efficiency and effectiveness gains.   

Our techniques are evaluated in terms of top $k$ recall at retrieval depth $d$, which we abbreviate as R@$(k, d)$.
For a given query vector $q$, the goal is to retrieve its top $k$ most similar vectors (in terms of cosine similarity), where we determine the ``ground truth'' by brute force (in the case of k--d trees, on the original vectors).
For each technique, Lucene can retrieve ranked results to any arbitrary depth $d$.
As is common, setting $d>k$ allows a refinement step (which we did not implement) where the {\it actual} similarity of all $d$ vectors can be computed and reranked to produced a final top-$k$ ranked list.

In our experiments, we set $k=10$, matching the application of using word embeddings in document ranking from Zuccon et al.~\cite{zuccon2015integrating}.
Specifically, we examined the settings $d = \{10, 20, 50, 100\}$.
The query terms used to perform evaluation were taken from the title of topics used in the TREC 2004 Robust Track, to match our retrieval application.
Latency measurements were performed on a common laptop (2.6 GHz Intel Core i7 CPU, 16GB of RAM; macOS 10.14.6; Oracle JDK 11)\ with $d=100$.
Note that query latency for top-$k$ retrieval in Lucene grows with $k$, and thus we are measuring the worst case setting.

\begin{table}[t]
\centering
\small
\setlength\tabcolsep{2pt}
\begin{tabular}{llrrrrrr}
\toprule
Model & Configuration & $d=10$ & $d=20$ & $d=50$ & $d=100$ & latency & index size\\
\toprule
{\bf word2vec} \\
fake words & $q=70$ & 0.64 & 0.82 & 0.94 & 0.97 & 234ms & 175MB\\
fake words & $q=60$ & 0.63 & 0.81 & 0.93 & 0.96 & 221ms & 190MB\\
fake words & $q=50$ & 0.62 & 0.81 & 0.92 & 0.96 & 209ms & 122MB\\
fake words & $q=40$ & 0.61 & 0.78 & 0.90 & 0.95 & 105ms & 96MB\\
fake words & $q=30$ & 0.57 & 0.74 & 0.87 & 0.93 & 97ms & 69MB\\
lexical LSH & $b=300, h=1, n=2$ & 0.51 & 0.65 & 0.79 & 0.85 & 193ms & 194MB\\
lexical LSH & $b=300, h=1, n=1$ & 0.55 & 0.72 & 0.84 & 0.91 & 245ms & 130MB\\
lexical LSH & $b=50, h=30, n=2$ & 0.51 & 0.65 & 0.79 & 0.85 & 196ms & 194MB\\
lexical LSH & $b=50, h=30, n=1$ & 0.55 & 0.72 & 0.84 & 0.91 & 276ms & 130MB\\
k--d tree & ppa-pca-ppa & 0 & 0.004 & 0.008 & 0.01 & 9ms & 14MB\\
k--d tree & pca & 0.008 & 0.01 & 0.02 & 0.03 & 11ms & 25MB\\
\midrule
{\bf GloVe} \\
fake words & $q=70$ & 0.64 & 0.83 & 0.95 & 0.98 & 220ms & 238MB\\
fake words & $q=60$ & 0.63 & 0.82 & 0.94 & 0.97 & 193ms & 202MB\\
fake words & $q=50$ & 0.62 & 0.81 & 0.93 & 0.97 & 225ms & 166MB\\
fake words & $q=40$ & 0.61 & 0.79 & 0.92 & 0.97 & 195ms & 167MB\\
fake words & $q=30$ & 0.57 & 0.75 & 0.89 & 0.94 & 132ms & 94MB\\
lexical LSH & $b=300, h=1, n=2$ & 0.50 & 0.65 & 0.80 & 0.87 & 176ms & 169MB\\ 
lexical LSH & $b=300, h=1, n=1$ & 0.51 & 0.70 & 0.85 & 0.91 & 196ms & 176MB\\
lexical LSH & $b=50, h=30, n=2$ & 0.50 & 0.65 & 0.80 & 0.87 & 203ms & 269MB\\
lexical LSH & $b=50, h=30, n=1$ & 0.52 & 0.72 & 0.84 & 0.91 & 278ms & 176MB\\
k--d tree & ppa-pca-ppa & 0.001 & 0.004 & 0.006 & 0.01 & 13ms & 19MB\\
k--d tree & pca & 0.002 & 0.002 & 0.006 & 0.01 & 16ms & 34MB\\
\bottomrule
\end{tabular}
\vspace{0.2cm}
\caption{R@(10,$d$) for different values of $d$, query latency, and index size. Parameters for the the various models are as follows:\ $q$ is the quantization factor for fake words; $b$ is the number of buckets, $n$ is the length $n$-grams, and $h$ is the number of hashes for lexical LSH; ppa-pca-ppa refers to Raunak~\cite{raunak2017simple} and pca refers to Wold et al.~\cite{wold1987principal}}
\label{tab:main}
\vspace{-0.5cm}
\end{table}

Our experimental results are presented in Table~\ref{tab:main}, where we show, for different approaches and settings, R@(10,$d$) for different values of $d$, as well as average query latency (at $d=100$) and index size.
It is clear that of the three approaches, fake words is the most effective as well as most efficient.
The k--d tree, while fast, yields terrible recall---the dimensionality reduction techniques discard too much information for the data structure to be useful.

\section{Conclusions}

It bears emphasis that Lucene was fundamentally not designed to support approximate nearest-neighbor search on dense vectors, and thus we are appropriating its indexing and retrieval pipeline for an unintended use.
As a result, our solutions lack elegance, but they do accomplish our goal of bringing approximate nearest-neighbor search into Lucene without any external dependencies.
For the builder of ``real-world'' search applications, the choice is between a single system (Lucene) that excels at retrieval with inverted indexes and imperfectly performs approximate nearest-neighbor search, or integrating a separate purpose-built system.
Ultimately, the decision needs to be considered within a broader set of tradeoffs, but the fake words approach might be compelling in certain scenarios.

\bibliographystyle{splncs04}

\end{document}